# Zero-shot Cross-domain Knowledge Distillation: A Case study on YouTube Music

Srivaths Ranganathan
Google LLC
Mountain View, USA
srivaths@google.com

Chieh Lo
Google LLC
New York, USA
chiehlo@google.com

Bernardo Cunha
Google LLC
New York, USA
becunha@google.com

Nikhil Khani
Google
Mountain View, USA
nkhani@google.com

Li Wei
Google
Mountain View, USA
liwei@google.com

Aniruddh Nath
Google
Mountain View, USA
aniruddhnath@google.com

Shawn Andrews
Google LLC
Mountain View, USA
shawnandrews@google.com

Gergo Varady
Google LLC
New York, USA
gvarady@google.com

Yanwei Song
Google LLC
Mountain View, USA
yanweisong@google.com

Jochen Klingenhoefer
Google LLC
New York, USA
jochenk@google.com

Tim Steele
Google LLC
New York, USA
timsteele@google.com

## Abstract

Knowledge Distillation (KD) has been widely used to improve the quality of latency sensitive models serving live traffic. However, applying KD in production recommender systems with low traffic is challenging: the limited amount of data restricts the teacher model size, and the cost of training a large dedicated teacher may not be justified. Cross-domain KD offers a cost-effective alternative by leveraging a teacher from a data-rich source domain, but introduces unique technical difficulties, as the features, user interfaces, and prediction tasks can significantly differ.

We present a case study of using zero-shot cross-domain KD for multi-task ranking models, transferring knowledge from a (100x) large-scale video recommendation platform (YouTube) to a music recommendation application with significantly lower traffic. We share offline and live experiment results and present findings evaluating different KD techniques in this setting across two ranking models on the music app. Our results demonstrate that zero-shot cross-domain KD is a practical and effective approach to improve the performance of ranking models on low traffic surfaces.

## CCS Concepts

• **Information systems** → **Recommender systems**; **Learning to rank**; • **Computing methodologies** → *Multi-task learning*.

## Keywords

Recommender Systems, Knowledge Distillation, Learning to Rank, Multitask Learning

## 1 Introduction

Modern recommender systems [3, 4] increasingly rely on large, complex models to deliver highly personalized user experiences. These models frequently reside on the critical path of serving recommendations, presenting a trade-off between prediction quality and inference latency. Knowledge Distillation (KD) [5, 6, 11] is a well-established technique enabling the transfer of knowledge from large, high-performing teacher models to smaller student models improving performance without a proportional increase in latency. Despite its successes, the effective application of KD in recommender systems with low traffic poses significant challenges [12]. The limited training data can restrict the feasible size of the teacher model: potentially leading to overfitting and diminishing returns. Furthermore, the operational overhead associated with training and maintaining a large dedicated teacher model may not be justifiable for applications with a smaller user base.

Cross-domain knowledge transfer [9] from a high-traffic source domain offers an alternative means to enhance model quality in smaller domains without increasing serving latency. Additionally, it can enable models to adapt to changing trends and new content which may appear more rapidly and in greater volume in the source domain.

However, realizing the gains from effective cross-domain knowledge transfer [2] introduces its own challenges, particularly when significant discrepancies exist between the source and target domains in terms of features, user interfaces or prediction tasks [1]. This paper addresses these challenges through a case-study of applying zero-shot [10] cross-domain KD to transfer knowledge from a large-scale video recommendation teacher model to student models serving traffic on a smaller scale Music recommender application.

## 2 Zero-shot Cross-Domain KD

To overcome the challenges of efficient knowledge transfer and resource optimization in our data-scarce platform, we propose leveraging an existing (zero-shot) shared "teacher" model [7], originally developed for YouTube, to improve ranking models on the YouTube Music application (cross-domain). Reflecting the lower scale of traffic to the music platform, the training data for the Music ranking models has 100x fewer samples than the data available from the YouTube app. We implemented an offline teacher label augmentation pipeline where the teacher model runs inference on the music dataset. The music ranking models were then modified to utilize these augmented teacher prediction labels during training. This approach yields considerable advantages, most notably substantial reductions in computational expenditure (GPU training hours) and minimal engineering maintenance overhead.

**Video Teacher Model:** The teacher model is a large-scale deep multi-task ranking model engineered to unify content discovery across multiple surfaces on YouTube. Its primary function is to generate "soft labels" for efficiently training more compact student models specific to various video recommendation surfaces on YouTube. The model architecture concatenates input features from disparate surfaces and processes them through multiple shared layers to extract a joint representation of the features relevant across multiple tasks. The shared representation then feeds into separate towers dedicated to distinct surface specific tasks.

However, leveraging this pre-trained Video teacher across domains and in a zero-shot manner for the Music application introduces several challenges stemming from their design and training data:

(1) **Feature set mismatch:** The Music ranking models and the Video teacher model evolved independently. This divergence results in a critical feature mismatch, where up to 40% of the input features expected by the teacher model are not extracted or available on the Music recommendation surfaces. Any features that are missing fallback to its default value during knowledge distillation.
(2) **Task and Label Distribution Divergence:** The objectives that the Video teacher model predicts may not directly align with the tasks on Music surfaces. Furthermore, even for similar tasks, the underlying label distributions can vary substantially between the video and music domains. For example, the YouTube Music Homepage displays content in grouped "shelves," unlike YouTube's feed of individual videos. This difference in presentation results in a click-through rate (CTR) variation of approximately 2% between the two surfaces.
(3) **User behavior patterns:** User consumption patterns for music differ markedly from those for videos. For instance, music engagement often involves more frequent repeat listening, lower exploration and longer sessions. These disparities lead to different underlying data distributions impacting the relevance of signals and interactions the teacher model was originally trained to recognize.

We successfully applied KD, despite the differences in the music and video domains, by distilling separate logits for teacher predictions (auxiliary distillation) to mitigate label biases, and picking distillation tasks such that the model learns a better representation of the input features in the shared layers, shared benefits across multiple tasks. We implemented our zero-shot cross-domain KD methodology on two distinct multi-task student ranking models on YouTube Music: the Homepage and the Music Radio rankers. Both models train with substantially less data and significantly smaller architectures (150x and 300x smaller for Homepage and Radio, respectively) compared to the large-scale YouTube video teacher model. Previously, both models also incorporated sampled data from the teacher's video domain for knowledge transfer in order to quickly adapt to new items and user behavior. Although this sampled-data approach introduced instability during model training and created dependencies on the non-target domain data, past attempts to remove this dependency were unsuccessful.

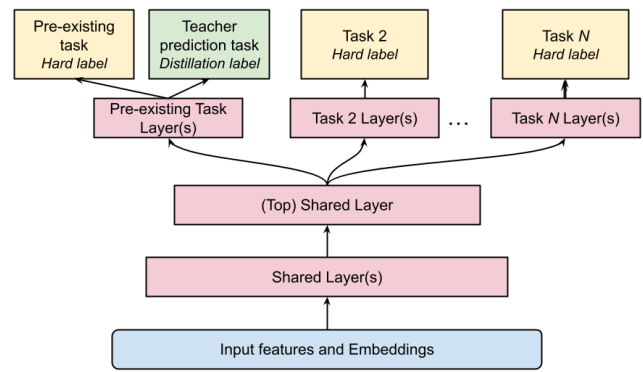

**Figure 1: Distillation for YouTube Music Homepage and Radio student models**

**Homepage model:** For the Music Homepage model, which predicts objectives such as CTR and trail engagement, we employed zero-shot auxiliary label distillation. The trail engagement task captures how long a user will listen conditioned on clicking the item. The model's towers for these two tasks were augmented to also predict the corresponding teacher labels from additional logits, allowing the model to train exclusively on the YouTube Music dataset. The Homepage model's third primary task, which predicts music discovery, was not distilled as it lacked a direct equivalent in the teacher model.

**Radio model:** The Music Radio ranker, responsible for sequencing music videos, presented a different scenario as it had no tasks in common with the video teacher model. To address this, we distilled a novel, "non-serving" task by adding a new prediction head to the student model. The new task trained to predict a soft label representing the teacher's "Continue Watching" task – an indicator of whether watching the current video would lead to subsequent video watches. This strategy allowed the Radio model to train solely on Music Radio data, effectively decoupling it from out-of-domain video sources and relying on KD for knowledge transfer.

## 3 Experiment Results

This section presents an empirical evaluation of our proposed zero-shot cross domain distillation setup. We apply the methodology to the two Music Multi-task ranking models described above comparing offline performance and online metrics against control models.

For our offline and online experiments, the baseline model shares the identical architecture to the student models, with respect to the number of layers and parameters for the shared layers and prediction towers, but lacking the auxiliary KD task heads and the corresponding losses. Note that the added auxiliary loss has the same weight as other losses. The control models were trained at the same time period as the student models to account for data freshness and number of trained steps.

### Finding 1: Zero-shot KD is effective despite lower teacher performance

Due to inherent differences in the UI and user behavior between the Video (teacher) and Music (student) applications, the teacher model has trained on a different underlying data distribution than that encountered on the Music surface. Consequently, for the specific tasks that were distilled, the Video-based teacher model has lower accuracy compared to the control model without distillation. Nevertheless, KD to the student model with the auxiliary task surpasses the baseline model in offline performance metrics: AUC for the CTR task and R-squared for the trail engagement regression task.

| Homepage Task | Control Model | Teacher Model | Cross-Domain Student Model |
|---|---|---|---|
| CTR (AUC) | 79.34 | 75.40 | **79.55** |
| Trail Engagement (R-squared) | 0.312 | 0.267 | **0.320** |

Table 1: Offline metrics for the Homepage student model

### Finding 2: Cross-task performance gains

We observe that applying KD to a subset of tasks within a student model enhances performance in other non-distilled tasks. For the Homepage model, we distill the CTR and the trail engagement task, but observe that the Discovery task also sees gains in offline performance. Furthermore, in the Radio model, a new non-serving task tower, distilled but not used on the Radio surface, contributes to increased AUC for the primary engagement task. This suggests that the cross-domain distillation improves the underlying shared representations, thereby conferring benefits across the various predicted tasks. Notably, for the Radio model, this is the only mechanism by which the online performance improves since we do not directly distill the tasks used on that surface.

| Task | Control AUC | Cross Domain Student AUC |
|---|---|---|
| Homepage Discovery Task | 76.06 | **76.22** |
| Radio Engagement Task | 90.30 | **91.38** |

Table 2: Offline metrics for non distilled tasks

### Finding 3: Zero-shot cross domain KD translates to significant online metric gains

Corroborating the offline gains detailed above, our live experiments validate the practicality and efficacy of this approach. Each experiment was run on YouTube Music for two weeks and we report statistically significant changes ($p<0.05$) in two primary metrics: an engagement metric and a Music discovery metric, the latter serving as an indicator of long term platform engagement. Given the scale of traffic to YouTube, newly released items occur much more frequently in the teacher model's dataset and consequently, it has better representation of these items. This results in an outsized impact on new releases we observed in both the experiments.

| Surface | Engagement metric | Discovery Metric | New Releases Engagement metric |
|---|---|---|---|
| Homepage | +0.58% | +1.12% | +11.39% |
| Radio | +0.70% | +2.13% | +0.96% |

Table 3: Online metrics for cross-domain KD students.

The discrepancy between the modest offline gains in the previous findings and the substantial online improvements below is a common phenomenon in industry recommender systems. The strong performance for the Discovery metric and the engagement with new releases suggests that the primary benefit of our KD approach is in improving a model's generalization and responsiveness to new content. Offline metrics like AUC cannot fully capture shifts in real world user behavior from increasing user satisfaction and novel content discovery, which drive overall platform health and promote long term engagement.

## 4 Conclusions And Future Work

In this work, we tackle the unique difficulties of improving recommender systems in low-traffic applications with strict latency constraints. We proposed and validated zero-shot cross-domain knowledge distillation (CDKD) as an effective strategy, showcasing its ability to transfer valuable knowledge from the data-rich YouTube video ecosystem to models serving the YouTube Music platform. Our paper presented techniques to apply distillation overcoming the gaps from differing tasks, feature spaces, and data characteristics between the source and target domains. Our real-world experiments, across two production ranking models on YouTube Music, offers compelling evidence that zero-shot CDKD can enhance student model performance even when the teacher model shows lower accuracy on the target domain, and that positive impacts can be observed on tasks not directly distilled. Furthermore, our results indicate that zero-shot CDKD is superior to directly training on source domain samples for generalizing to new items and adapting to evolving content trends.

Looking ahead, we aim to apply this methodology to other models and surfaces within YouTube Music. We also plan to perform more comprehensive analysis e.g., ablating the teacher's predictions with random noise to analyze the impact of the privileged information [8]. We also plan to investigate the impact of CDKD as the student model's feature set becomes more comprehensive and as the student models themselves are scaled, to fully unlock the potential of this promising technique.

## 5 Speaker Bio

**Srivaths Ranganathan** is a Staff Software Engineer at YouTube (Google), where he works on improving YouTube Music's ranking algorithms.

**Chieh Lo** is a Senior Software Engineer at YouTube (Google), where he works on improving YouTube Music's Radio Ranking.

**Bernardo Cunha** is a Senior Software Engineer at YouTube (Google), where he works on improving YouTube Music's ranking algorithms.